# Projection Decomposition for Dual-energy Computed Tomography

Wenxiang Cong, Daniel Harrison, Yan Xi, Ge Wang

*Abstract*—Dual-energy computed tomography (CT) is to reconstruct images of an object from two projection datasets generated from two distinct x-ray source energy spectra. It can provide more accurate attenuation quantification than conventional CT with a single x-ray energy spectrum. In the diagnostic energy range, x-ray energy-dependent attenuation can be approximated as a linear combination of photoelectric absorption and Compton scattering. Hence, two physical components of x-ray attenuation can be determined from two spectrally informative projection datasets to achieve monochromatic imaging and material decomposition. In this paper, a projection decomposition method is proposed for the image reconstruction in dual-energy CT. This method combines both an analytical algorithm and a single-variable optimization method to solve the non-linear polychromatic x-ray integral model, allowing accurate quantification of photoelectric absorption and Compton scattering components. Numerical tests are performed to illustrate the merit of the proposed method by comparing with classical projection decomposition methods.

## I. Introduction

COMPUTED tomography (CT) can reconstruct a three-dimensional image of an object from a series of projections, providing important diagnosis information. In clinical CT, an x-ray source is polychromatic, and x-ray detectors are currently operated in a current-integrating mode. CT image reconstruction is based on an approximate line integral model, ignoring x-ray energy information. However, lower energy photons are more easily absorbed than higher energy photons, which would cause the x-ray beam to become increasingly harder as it propagates through the object [1]. This physical model mismatch would generate significant beam-hardening artifacts in the reconstructed image. Dual-energy CT is a well-established technique, allowing monochromatic imaging and material decomposition [2, 3]. Current dual-energy x-ray imaging can be implemented by kVp-switching, dual-layer detection, dual-source scanning, and simplistic two-pass scanning.

This work was supported by the National Institutes of Health Grant NIH/NIBIB R01 EB016977 and U01 EB017140.

W. Cong D. Harrison, and G. Wang are with the Biomedical Imaging Center, Department of Biomedical Engineering, Rensselaer Polytechnic Institute, Troy, NY 12180 USA (congw@rpi.edu, danharrison@nycap.rr.com, wangg6@rpi.edu).

Y. Xi is with First Imaging Technology, Shanghai 201318, China (xiyansjtu@gmail.com).

Several image reconstruction methods for dual-energy CT were developed over the years. Alvarez and Macovski proposed an image reconstruction method in the projection domain by solving a non-linear integral equation to decompose dual-energy measurements into two independent sinograms, each of which corresponds to a basis component [2]. The lookup table method was designed for projection decomposition with a table mapping between dual-energy x-ray intensity measurements and basis material projections according to a polychromatic nonlinear integral model. Then, projections of basis material coefficients can be calculated from measured dual-energy data via interpolation of data in the lookup table [4]. However, the lookup table method is limited by storage size and interpolation accuracy. Alternatively, image-domain reconstruction methods first reconstruct images from the low- and high-energy sinograms using filtered back projection (FBP), and then perform image-domain material decomposition [5, 6]. This type of image-domain reconstruction makes substantial approximations in energy spectra, resulting in quantitatively inaccurate results [7]. Recently, iterative methods incorporate an accurate physical model to reconstruct images directly from dual-energy measurements [8]. These approaches involve a highly nonlinear forward model in the maximum likelihood framework to model the polychromatic measurement, representing a complicated nonlinear optimization problem. The great computation cost and slow convergence speed significantly reduces the practicality of the algorithm.

In this paper, a new image reconstruction approach for dual-energy CT is proposed based on a realistic polychromatic physical model. This method combines an analytical algorithm and a single-variable optimization method to solve the non-linear polychromatic x-ray integral model in the projection domain, allowing an efficient and accurate decomposition for sinograms of two physical basis components. In the next section, the physical model and reconstruction methods are described. In the third section, representative numerical experiments are presented. In the last section, relevant issues are discussed.

## II. Methodology

An x-ray source in CT generally emits a polychromatic spectrum of x-ray photons, and the x-ray linear attenuation through the object depends on the object material composition and the photon energy. After a x-ray beam passes through the object, the x-ray intensity $I$ measured by a current-integrating detector can be described by the non-linear integral model [2]:

$$I = \int_{E_{\min}}^{E_{\max}} S(E)D(E)\exp\left(-\int_l \mu(r,E)dr\right)dE, \quad (1)$$

where $S(E)$ is the energy distribution (spectrum) of the x-ray source, $D(E)$ is the detection efficiency, and $\mu(r,E)$ is the linear attenuation coefficient at an energy $E$ and a spatial position $r$ along a linear path $l$ through the object. During propagation through the object, the x-ray photons population is statistically attenuated according to the nonlinear equation (1).

It is well known that photoelectric absorption and Compton scattering are the two dominant x-ray attenuation processes in the 20 keV-140 keV diagnostic energy range [2]. The resulting x-ray linear attenuation coefficient can be represented by [2, 9, 10]:

$$\mu(r,E) = \rho \frac{N_A}{A}\left(\sigma_{ph} + \sigma_{co}\right), \quad (2)$$

where $\rho$, $N_A$, and $A$ are mass density, Avogadro's number ($6.022 \times 10^{23}$ atom/g-atom) and atomic mass respectively. The photoelectric atomic cross section, $\sigma_{ph}$, is formulated as [11]

$$Z^4 \alpha^4 \frac{8}{3}\pi r_e^2 \sqrt{\frac{32}{\varepsilon^7}} \quad \text{for } \varepsilon < 1, \quad (2a)$$

where $\varepsilon = E/511\,\text{keV}$, $Z$ is the atomic number, $\alpha$ is the fine-structure constant ($\approx 1/137$), and $r_e = 2.818$ fm is the classical radius of an electron. The Compton atomic cross section $\sigma_{co}$ is formulated as $Z f_{kn}$, here $f_{kn}$ is the Klein-Nishina function:

$$f_{kn}(\varepsilon) = 2\pi r_e^2 \left( \frac{1+\varepsilon}{\varepsilon^2}\left[\frac{2(1+\varepsilon)}{1+2\varepsilon} - \frac{1}{\varepsilon}\ln(1+2\varepsilon)\right] \right.$$
$$\left. + \frac{1}{2\varepsilon}\ln(1+2\varepsilon) - \frac{1+3\varepsilon}{(1+2\varepsilon)^2} \right) \quad (2b)$$

With both photoelectric and Compton atomic cross sections, the associated linear attenuation coefficients can be expressed as the product of spatial-dependent and energy-dependent components:

$$\mu(r,\varepsilon) = a(r)p(\varepsilon) + c(r)q(\varepsilon), \quad (3)$$

where

$$a(r) = \rho Z^4 / A \quad (3a)$$

is the spatial-dependent photoelectric component,

$$c(r) = \rho Z / A \quad (3b)$$

is the spatial-dependent Compton scattering component,

$$p(\varepsilon) = N_A \alpha^4 \frac{8}{3}\pi r_e^2 \sqrt{\frac{32}{\varepsilon^7}} \quad (3c)$$

is the energy-dependent photoelectric component, and

$$q(\varepsilon) = N_A f_{kn}(\varepsilon) \quad (3d)$$

is the energy-dependent Compton scattering component. For compound matter, the mass density $\rho$, atomic number $Z$, and atomic mass $A$ should be the effective density and effective atomic number and effective atomic mass respectively [12].

With dual-energy CT, we have two distinct x-ray intensity measurements $I_1$ and $I_2$ at each projection angle associated with a given x-ray source spectrum:

$$\begin{cases} I_1 = \int_{\varepsilon_{\min}}^{\varepsilon_{\max}} S_1(\varepsilon)\exp\left(-p(\varepsilon)\int_l a(r)dr - q(\varepsilon)\int_l c(r)dr\right)d\varepsilon \\ I_2 = \int_{\varepsilon_{\min}}^{\varepsilon_{\max}} S_2(\varepsilon)\exp\left(-p(\varepsilon)\int_l a(r)dr - q(\varepsilon)\int_l c(r)dr\right)d\varepsilon \end{cases} \quad (4)$$

The projection decomposition in dual-energy CT is to compute the projections $\int_l a(r)dr$ and $\int_l c(r)dr$ of spatial-dependent photoelectric absorption and Compton scattering components from x-ray dual-energy intensity measurements $I_1$ and $I_2$. Using the first x-ray energy spectral measurement, we have,

$$I_1 = \int_{\varepsilon_{\min}}^{\varepsilon_{\max}} S_1(\varepsilon)\exp\left(-p(\varepsilon)\int_l a(r)dr - q(\varepsilon)\int_l \bar{c}(r)dr\right)$$
$$\times \exp\left(-q(\varepsilon)\int_l \left[c(r) - \bar{c}(r)\right]dr\right)d\varepsilon \quad (5)$$

The second exponential term in Eq. (5) is expanded into a fourth-order Taylor polynomial,

$$\begin{cases} \exp(-q(\varepsilon)x) \approx 1 - q(\varepsilon)x + \frac{1}{2}q^2(\varepsilon)x^2 - \frac{1}{6}q^3(\varepsilon)x^3 + \frac{1}{24}q^4(\varepsilon)x^4 \\ x = \int_l \left[c(r) - \bar{c}(r)\right]dr \end{cases} \quad (6)$$

where $\bar{c}(r)$ is an initial estimation of the spatial-dependent Compton scattering component $c(r)$. The fourth-order polynomial approximation has a high accuracy with the use of the initial estimation $\bar{c}(r)$. For example, the mass density, atomic mass and atomic number of water may be used as initial estimates for biomedical dual-energy CT. Also, the initial estimation can be obtained using the low resolution lookup table method. Generally, the exponential of the second exponential term in Eq. (5) satisfies: $\left|q(\varepsilon)\int_l \left[c(r) - \bar{c}(r)\right]dr\right| < 1.1$ for human tissues with bone volume fraction of 12.9% or less [13]. The error of the fourth-order Taylor approximation will not exceed 1.5% under the following conditions: bone density 1.92g/cm$^3$, Z/A of bone is 0.51478, muscle density 1.0599g/cm$^3$, Z/A of muscle 0.55, and the x-ray path length through the human body 35cm.

Inserting Eq. (6) into Eq. (5) and performing integral in term of energy variable $\varepsilon$, we have

$$\begin{cases} I_1 = \int_{\varepsilon_{\min}}^{\varepsilon_{\max}} S_1(\varepsilon)\exp\left(-p(\varepsilon)\int_l a(r)dr - q(\varepsilon)\int_l \bar{c}(r)dr\right) \times \\ \qquad \left[1 - q(\varepsilon)x + \frac{1}{2}q^2(\varepsilon)x^2 - \frac{1}{6}q^3(\varepsilon)x^3 + \frac{1}{24}q^4(\varepsilon)x^4\right]d\varepsilon \\ \quad = \left[p_0(y) + p_1(y)x + p_2(y)x^2 + p_3(y)x^3 + p_4(y)x^4\right], \\ y = \int_l a(r)dr \end{cases} \quad (7)$$

Eq. (7) is a quartic algebraic equation, and there are analytic solutions. Because the right-hand side of Eq. (5) is an exponential function with respect to the variable $x$, the polynomial function is strictly convex, yielding two real roots and a pair of conjugate complex roots. From the prior of value range of $x$, one can identify the true solution, denoted as $x = h(y)$.

Furthermore, applying the second spectral measurement, the projection of the spatial-dependent photoelectric absorption distribution can be computed from the following single variable optimization,

$$\begin{cases} y_{opt} = \arg\min \left\| I_2 - \int_{\varepsilon_{\min}}^{\varepsilon_{\max}} S_2(\varepsilon)\exp\left[-p(\varepsilon)y - q(\varepsilon)h(y)\right]d\varepsilon \right\| \\ y_{opt} = \int_l a(r)dr \end{cases} \quad (8)$$

Eq. (8) can be effectively solved via single variable optimization, such as golden section search and parabolic interpolation. Therefore, the projections of spatial-dependent photoelectric absorption and Compton scattering images can be effectively determined by solving Eqs. (7) and (8) simultaneously for every detector elements at each projection view. Given a major involvement of these analytic steps, the proposed algorithm for projection decomposition is called a quasi-analytic method.

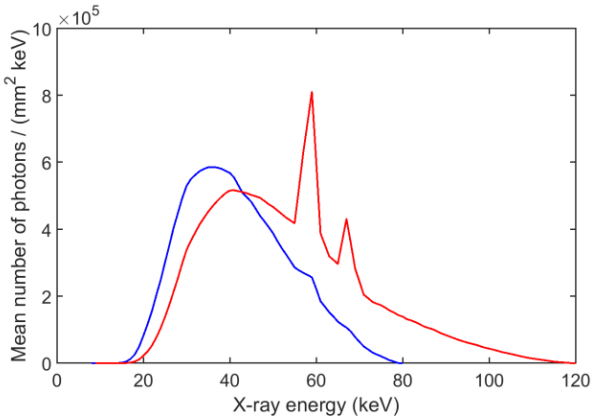

Fig. 1. Energy spectral distributions of x-ray source simulated using the public software from Siemens website (https://health.siemens.com/booneweb/index.html). (a) The energy spectrum generated from the x-ray tube (80 kVp) filtered by Aluminum of 2mm thickness, and (b) the energy spectrum from the x-ray tube (120 kVp) filtered by Aluminum of 3mm thickness.

## III. NUMERICAL EXPERIMENTS

In the numerical simulation, two x-ray energy spectra were respectively generated from x-ray tube operated at 80 kVp/20mA and 140 kVp/20mA to simulate dual-energy imaging, as shown in Fig. 1. A numerical phantom was designed to evaluate methods of the projection decomposition. The circular phantom has a diameter of 44cm, and contained 9 sub-regions that were filled with various materials to simulate human tissues. The effective atomic numbers, densities, and atomic masses in these sub-regions are listed in Table 1, which characterized photoelectric and Compton cross-sections, as shown in Fig. 2. The phantom was placed at isocenter. The source-to-isocenter distance was set to 54.1 cm and the source-to-detector distance was set to 94.9 cm for the x-ray dual-energy imaging.

The phantom was discretized into 512×512 square pixels. The energy-dependent linear attenuation coefficients were synthesized according to Eq. (3). Based on two energy spectra shown in Fig. 1, the dual-energy projection datasets were generated based on Eq. (1) for 180 views over a range of 180°. Then, the projection data were corrupted by Poisson noise to simulate real experiments.

**Table 1. Parameters of the numerical phantom.**

| Tissue | 1 | 2 | 3 | 4 | 5 | 6 | 7 | 8 | 9 |
|---|---|---|---|---|---|---|---|---|---|
| Z | 3.04 | 4.90 | 4.02 | 5.15 | 5.09 | 4.58 | 5.91 | 5.39 | 5.64 |
| P | 1.00 | 0.96 | 0.99 | 1.07 | 1.06 | 1.03 | 1.20 | 1.06 | 1.10 |
| A | 6.49 | 8.93 | 7.35 | 8.90 | 9.34 | 13.96 | 8.22 | 10.12 | 9.32 |

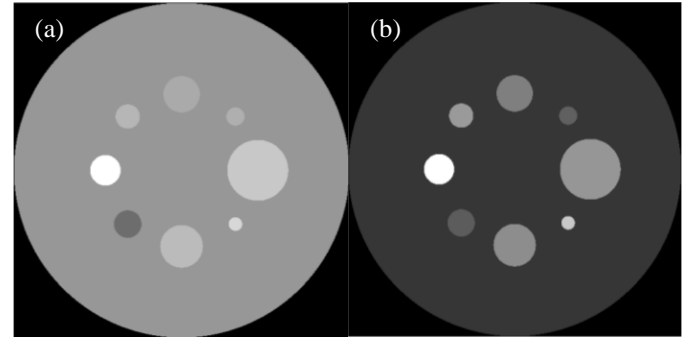

Fig. 2. Numerical phantom. (a) The true Compton scattering image, and (b) the true photoelectric absorption image.

**Quasi-analytic method:** The MATLAB programming was developed to implement the proposed algorithm for the projection decomposition. The projection of spatial-dependent Compton scattering component and projection of photoelectric component were computed from the dual-energy projection dataset of the phantom. Then image reconstructions were performed to obtain the photoelectric-absorption and Compton-scattering images using the classical filtered backprojection (FBP) method. The reconstructed Compton scattering image and photoelectric absorption image had a good spatial and contrast resolution, and were in excellent agreement with the truth, as shown in Fig. 3(a-d). The detailed features of the reconstructed images are quantitatively accurate, and the beam hardening effect is insignificant. Furthermore, the attenuation coefficient in each energy bin can be computed from the reconstructed Compton scattering and photoelectric absorption images based on Eq. (3) to synthesize a monochromatic image reconstruction or enable material decomposition. The computational program took about 10 minutes in a PC with 16G RAM and 2.8GHz CPU.

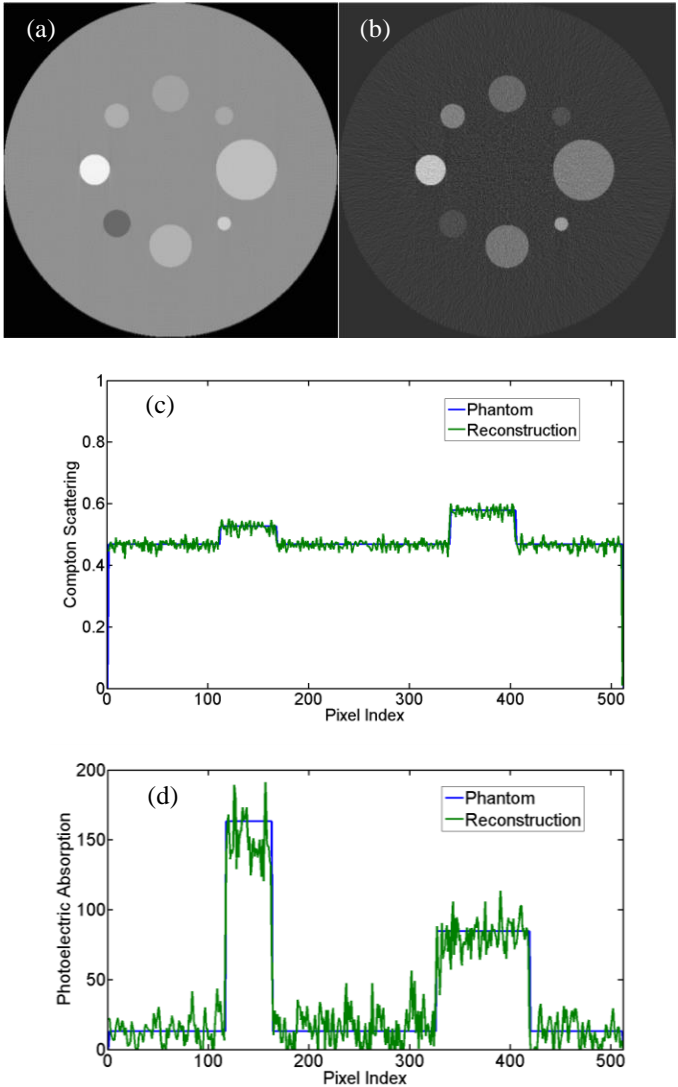

Fig. 3. Image reconstructions via the quasi-analytic method. (a) The reconstructed Compton scattering image, (b) the reconstructed photoelectric absorption image, (c) the profiles along the vertical midlines in the phantom and reconstructed Compton scattering images, and (d) the profiles along the horizotal midlines in the phantom and reconstructed photoelectric absorption images respectively.

For comparison with the same dataset and computing environment, classical projection decomposition methods, including the lookup table method and the direct optimization method were also implemented to illustrate the merit of our proposed method.

**Lookup table method:** We first established a mesh grid for unknown variables $(x, y) = \left( \int_l a(r) dr, \int_l c(r) dr \right)$ to generate a solution set:

$$S = \left\{ (x_i, y_j) \mid x_i = i\Delta_x; y_j = j\Delta_y; i = 1, 2, \cdots, N_x; j = 1, 2, \cdots, N_y \right\}$$

with $\Delta_x = 0.05$, $\Delta_y = 0.04$, $N_x = 60000$, and $N_y = 1000$. From the solution set, we generated x-ray intensity datasets $\{I_1(x_i, y_j), I_2(x_i, y_j)\}$ based on Eq. (4). Then, we obtained an inverse functional relationship $f(I_1, I_2)$ from the datasets $(I_1, I_2)$ to projections of photoelectric absorption and Compton scattering components $(x, y)$. Then, the inverse function was applied to perform projection decomposition via interpolation. Finally, the images were reconstructed using FBP to obtain the Compton-scattering and photoelectric-absorption images respectively, as shown in Fig. 4. However, the lookup table method produced noisy photoelectric-absorption image since $p(\varepsilon)$ is significantly smaller than $q(\varepsilon)$, having a deminishing gradient with respect to variable $y$. This method also requires a huge size of memory, and takes a high computational cost. In the example, the algorithm took up 10GB memory and took about 62 minutes in the same computer environment.

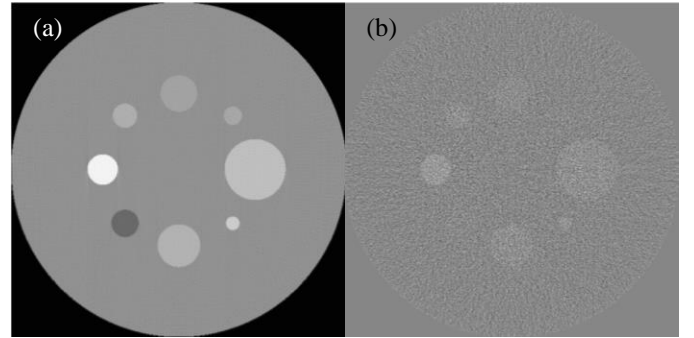

Fig. 4. Image reconstructions using the lookup table method. (a) The reconstructed Compton scattering image, and (b) the reconstructed photoelectric absorption image.

**Direct optimization method:** This method is to solve Eq. (4) through direct optimization. The nonlinear optimization process was implemented using the MatLab optimization function fminsearch(), which needs initial values. Due to measurement noise, the optimization algorithm will converge prematurely to an inaccurate solution if the initial guess is not close enough to the true solution. In the numerical computation, we observed many cases that our quasi-analytic method outperformed the direct optimizaiton method when the latter method was trapped in local minima. For example, if we used the direct optimization method from zero initial values to perform the projection decomposition. The direct optimization method produced very noisy photoelectric-absorption image, as shown in Fig.5.

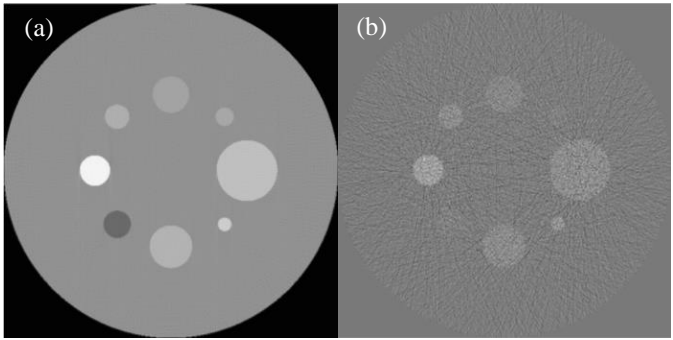

Fig. 5. Image reconstructions through the direct optimization method from zero initial values. (a) The reconstructed Compton scattering image, and (b) the reconstructed photoelectric absorption image,

Then, as the best senario for the direct optimization method, we used the result obtained using the lookup table method as initial values to perform the projection decomposition. From these initial values, the direct optimization method gave Compton scattering and photoelectric absorption images similar to that obtained using the quasi-analytic method, as shown in Fig. 6. However, its computational cost was more than six times higher than that of the quasi-analytic method, and yet it utilized the result obtained using the lookup table method that took a large amount of memory.

## IV. Conclusion

Dual-energy computed tomography (CT) is to reconstruct images of an object from two projection datasets obtained from two distinct x-ray source spectra, allowing more accurate quantification of the attenuation section/volume. In the diagnostic energy range, the x-ray attenuation is essentially a combination of the photoelectric and Compton scattering effects. The interactions are energy-dependent. Thus, measurements at two distinct energies should permit the separation of the attenuation into its basic components, providing both density and atomic number information of involved materials.

The projection decomposition for dual-energy CT is a complicated nonlinear inverse problem. A direct analytical solution does not exist, and a direct optimization method is often used to solve the bivariate nonlinear integral equation system. In this study, a fast and effective method has been proposed to solve the system of two non-linear polychromatic x-ray integral equations. This method combines an analytical algorithm and a single-variable optimization method for low computational cost and accurate projection decomposition, realizing monochromatic imaging and material decomposition and avoiding the beam hardening problem in conventional CT. Numerical results have shown that the proposed image reconstruction method has made improvements comparing with classical projection decomposition methods. Currently, we are seeking real datasets to further evaluate the image quality in clinical applications. The proposed method is also applicable to nondestructive testing and security screening.

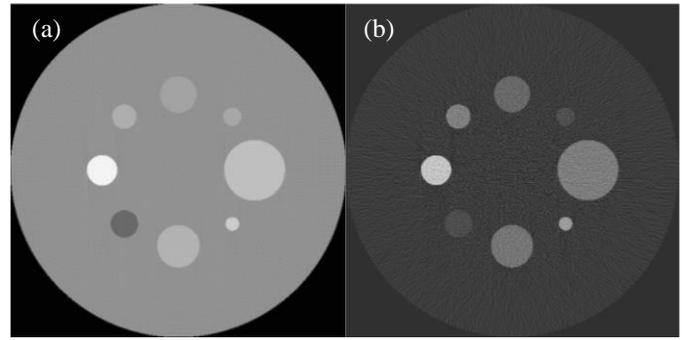

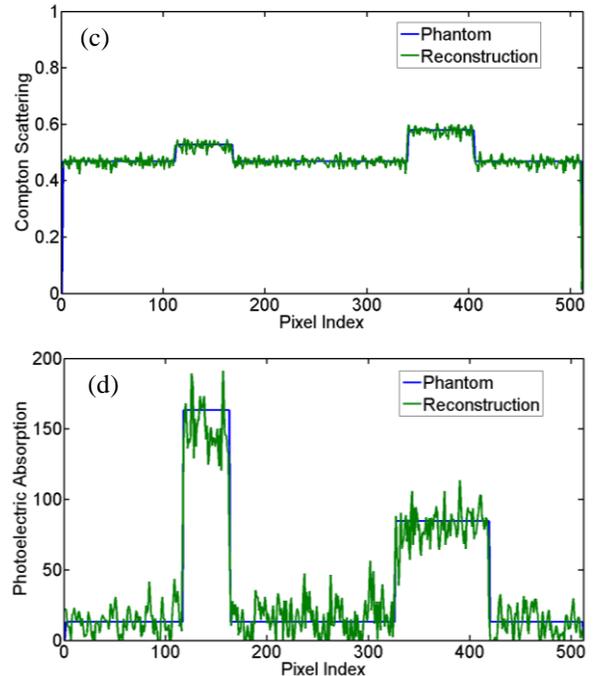

Fig. 6. Image reconstructions through the direct optimization method. (a) The reconstructed Compton scattering image, and (b) the reconstructed photoelectric absorption image, (c) the profiles along the vertical midlines in the phantom and reconstructed Compton scattering images, and (d) the profiles along the horizotal midlines in the phantom and reconstructed photoelectric absorption images respectively.

Acknowledgment: This work is partially supported by the National Institutes of Health Grants NIH/NIBIB R01 EB016977 and U01 EB017140.